\documentclass[%
twocolumn,
superscriptaddress,
 amsmath,amssymb,
 aps,
pra,
floatfix
]{revtex4}

\usepackage{subfigure}

\usepackage{color}

\usepackage{graphicx}
\usepackage{bm}
\usepackage{amsmath}
\usepackage{amssymb}

\newtheorem{remark}{Remark}

\begin{document}

\title{Error-tolerant oblivious transfer in the noisy-storage model}
\author{Cosmo Lupo}
\affiliation{Dipartimento  Interateneo di Fisica, Politecnico di Bari \& Universit\`a di Bari, 70126, Bari, Italy}
\affiliation{INFN, Sezione di Bari, 70126 Bari, Italy}
\author{James T. Peat}
\author{Erika Andersson}
\affiliation{SUPA, Institute of Photonics and Quantum Sciences, School of Engineering and Physical Sciences,
Heriot-Watt University, EH14 4AS Edinburgh, United Kingdom}
\author{Pieter Kok}
\affiliation{Department of Physics and Astronomy, The University of Sheffield, S3 7RH Sheffield, United Kingdom}

\date{\today}

\begin{abstract}
The noisy-storage model of quantum cryptography allows for information-theoretically secure two-party computation based on the assumption that a cheating user has at most access to an imperfect, noisy quantum memory, whereas the honest users do not need a quantum memory at all.
In general, the more noisy the quantum memory of the cheating user, the more secure the implementation of oblivious transfer, 
which is a primitive that allows universal secure two-party and multi-party computation.
For experimental implementations of oblivious transfer, one has to consider that also the devices held by the honest users are lossy and noisy, and error correction needs to be applied to correct these trusted errors.
The latter are expected to reduce the security of the protocol, since a cheating user may hide themselves in the trusted noise.
Here we leverage entropic uncertainty relations to derive tight 
bounds on the security of oblivious transfer with a
trusted and untrusted noise.
In particular, we discuss noisy storage and bounded storage, with independent and correlated noise.
\end{abstract}

\maketitle


\section{Introduction}

Two-party computation denotes a family of problems whose goal is to allow two users, Alice and Bob, who do not trust each other, to evaluate a function of two arguments, $f(x,y)$, where Alice provides $x$ and Bob provides $y$.
Informally, the protocol is secure if no more information leaks to Alice about $y$, or to Bob about $x$, than what they can infer from the value of $f(x,y)$.
For example, $x$ and $y$ are $n$-bit strings and $f(x,y) = x \cdot y$, where $\, \cdot \,$ is the scalar product modulo 2. Neither Alice nor Bob will always be able to infer both $x$ and $y$ from $f(x,y)$ and their respective inputs $x$ or $y$.
In a highly influential paper \cite{Kilian}, Kilian showed that the ability to perform  
Oblivious Transfer (OT) securely is sufficient for two-party computation.
In OT, a receiver (Bob) of two incoming bits sent by Alice can read out exactly one of them, thereby making the readout of the other bit impossible.
The protocol is secure if Bob obtains no information about the other bit, and if Alice does not know which of the two bits is learnt by Bob.
Specifically, this is known as 1-out-of-2 OT (1-2 OT) \cite{12OT}. 
While OT has been formulated in a number of different flavours 
\cite{Rabin,Brassard,OT}, 
we focus on 
~1-out-of-2 randomised oblivious string transfer (1-2 $\text{ROT}^\ell$).
Other important primitives are Bit Commitment (BC) and Weak String Erasure (WSE). Both OT and BC can be obtained from WSE~\cite{Konig}.

Implementations of two-party computation with computational security leverages the complexity of solving some hard mathematical problems, e.g.~factoring \cite{Kilian,Konig}. 
Quantum mechanics cannot help in making two-party computation unconditionally secure \cite{Mayers,Lo}.
However, information-theoretic security can be achieved if the users have constrained capabilities, for example if they have access only to a limited pool of quantum resources.
This may happen if the users have no quantum memory \cite{Buhrman,DiVincenzo,Guha}, or if they have an imperfect quantum memory that can store only a limited number of qubits, for a limited time \cite{Entropy,Quantum}, or with non-unit fidelity.
The assumption of a quantum memory with limited capacity is known as the {\it bounded quantum storage model} \cite{OT,bounded,secureID}.
In general, the memory can be bounded and noisy, in which case one refers to the {\it noisy-storage model} \cite{Noise-0,Noise,Konig,CSPRA}.
Protocols that achieve OT, BC, and WSE within the noisy-storage model have been discussed in detail in Refs.~\cite{OT, Konig, bounded, secureID, Noise-0, Noise, Wehner-Lo, CSPRA, ES, NGPRA} and demonstrated experimentally~\cite{Ng, Erven}.
In general, experimental implementations of two-party computation are more demanding than Quantum Key Distribution (QKD). 
This is essentially due to the fact that users of two-party computation do not trust each other,
which places limits on how they can cooperate.
However, Refs.~\cite{OT, Konig, bounded, secureID, Noise-0, Noise, Wehner-Lo, CSPRA, ES, NGPRA,Ng, Erven}
have shown that primitives such as OT and BC are nevertheless feasible and secure within the noisy-storage model.
Device-independent OT can also be achieved by invoking suitable assumptions~\cite{Ribeiro,AB}.
Otherwise, without imposing any constraint on the users' resources, two-party computation can be implemented with partial security. 
In this latter case, one aims at computing bounds on the probability of successful cheating of a dishonest user, see e.g.~Refs.~\cite{Amiri,Stroh} and references therein.

In this paper we analyse the security of OT within the noisy-storage model.
We focus on the protocol of Damgård \textit{et al.}~\cite{OT}, which in turn is based on the same building blocks as the well-known BB84 protocol of QKD \cite{BB84}.
Building on previous results from Refs.~\cite{Konig,ES}, we derive tighter entropic uncertainty relations and quantify the security of OT in terms of the conditional min-entropy.
Compared to previous works, we extend the region of experimental parameters that allows for secure OT, and improve the trade-off between trusted noise (from the devices of the honest users) and noise in the quantum memory of a cheating user.
We show that, in order to achieve secure OT, the overall trusted noise should be below $\simeq 22\%$, for a noisy but unbounded quantum memory, and that this value decreases with decreasing noise in the quantum memory.

The paper is organised as follows.
We start in Section~\ref{Sec:OT} by recalling the implementation of 1-2 $\text{ROT}^\ell$ of Ref.~\cite{OT}, including measures to make it robust to loss and noise~\cite{Noise-0,Noise,Lo}.
The security of this protocol in the noisy-storage model is discussed in Section \ref{Sec:sec}.
In Section \ref{Sec:entropic} we review the entropic uncertainty relations derived in Refs.~\cite{OT,Konig,ES} and their application to prove the security of OT.
We leverage these results to obtain new, tighter entropic bounds in Section~\ref{Sec:improved}.
In Section~\ref{Sec:tolerant}, these new bounds are applied to characterise the security of noise-resilient OT and to determine the trade-off between trusted and untrusted noise.
In Section~\ref{Sec:corr} we discuss a quantum memory with correlated noise, showing that the new bounds are particularly advantageous in this case.
Finally, conclusions are presented in Section~\ref{Sec:end}.


\section{Oblivious transfer}\label{Sec:OT}

In this paper we focus on the task of 1-2 $\text{ROT}^\ell$, i.e.~1-out-of-2 randomised oblivious string transfer, and on the protocol introduced by Damgård \textit{et al.}~in Ref.~\cite{OT}.
In 1-2 $\text{ROT}^\ell$, the sender Alice outputs two random strings, each of $\ell$ bits. 
The receiver Bob outputs only one of these strings, chosen at random.
The task is executed securely if: 
(sender security) Bob gets little or no information about the other string, and
(receiver security) Alice does not know which string has been obtained by Bob.
The task can be realised using BB84 operations \cite{OT}, where Alice randomly prepares states using two conjugate bases, and Bob independently measures in either basis. Bob makes a single choice of basis in which he measures all the physical bits received.

Furthermore, the protocol can be made resilient to noise by adding a layer of error correction~\cite{Noise-0}.
We need to use some care when dealing with error correction in two-party computation since, in contrast to QKD, here Alice and Bob do not trust each other. Therefore, they cannot cooperate to estimate the loss and noise in the communication channel and in their devices, as they would do in QKD.
\begin{remark}
To implement error correction in two-party computation, the loss and noise need to be well characterised before running the protocol. This includes errors in the communication channels and in the devices held by the honest users. 
\end{remark}
This is a non-trivial requirement. Note that if the users exploit a trusted third party to certify the level of noise, then they can also implement OT using said trusted third party with a classical protocol.

In view of photonic applications, and following Refs.~\cite{Noise-0,Noise,Lo}, it makes sense to treat loss and noise in two different ways.
To deal with photon loss, Alice and Bob first need to agree on a common time reference, which allows them to time-tag each photon sent by Alice. 
If Bob is honest, he will measure each photon as soon as he gets them, and will confirm receipt to Alice.
Only the photons that made it to Bob will be actually used for the protocol. 
All the other, lost photons, will be disregarded \footnote{This is a nontrivial assumption, as it may generally allow a dishonest receiver to cheat more often by claiming that they did not receive the photon when they failed in cheating in some way, and also may allow a sender to cheat, by sending pulses that are more or less likely to be lost; the fact whether the pulse is lost or not can provide information on how it was measured, or what the result was more likely to be (see Ref.~\cite{Damian} discussing related issues). In this work, this issue is not considered because we assume that loss and noise are well characterised (see Section \ref{Sec:sec}).}. 
Photons that are eventually detected by Bob may still be subject to noise in their internal degrees of freedom. 
To deal with this noise, Alice will send an error syndrome to allow Bob to error-correct.

The error-tolerant protocols is as follows~\cite{Noise-0}.
\begin{itemize}

\item Alice randomly chooses the binary values $x'_j \in \{ 0, 1 \}$ and $\theta'_j \in \{ 0 , 1 \}$, for $j=1,2,\dots,n'$. According to the agreed time reference, at the $j$th time Alice prepares and sends to Bob the state $|x'_j,\theta'_j\rangle = H^{\theta'_j} |x'_j\rangle$, where $|x'_j\rangle \in \{ |0\rangle, |1\rangle \}$ is an element of the computational basis, and $H$ is the Hadamard gate.

\item Bob randomly chooses a bit value $B \in \{ 0 , 1 \}$, and measures all the qubits he receives in the basis $\{ H^B |0\rangle, H^B |1\rangle \}$.

\item Due to photon loss and detection inefficiency, some of the qubits are erased.
Bob keeps track of the timings when no photon is detected, and communicates this information to Alice.
Only the bit values associated to the photons detected are retained for the rest of the protocol.
If $n$ photons are detected, this identifies the sub-strings $X=x_1,\dots, x_n$ and $\Theta = \theta_1,\dots,\theta_n$ associated to Alice's binary values and basis choices. On Bob's side, he keeps track of the bit values measured in the basis of his choice, $Y=y_1,\dots,y_n$.

\item The overall loss is described by an attenuation factor $\eta \in (0,1)$, which is the probability that a photon is detected by Bob. As stated above, the honest users know the expected value of this parameter. The protocol aborts if such a value is not compatible with the empirical loss factor $n/n'$ and, within statistical errors, independent of the state prepared by Alice.

\item The protocol pauses for a waiting time $\Delta t$, counted from when Bob is expected to receive the last photon.

\item Alice announces her basis choices.
This corresponds to revealing two subsets of indices $\mathcal{I}_0$, $\mathcal{I}_1$, where
$\mathcal{I}_C = \{ j \, | \, \theta_j = C \}$.
She also announces a pair of hash functions $F_0$ and $F_1$ from $n$ to $\ell$, and the syndrome vectors
$\Sigma_0 = \text{syn}(X_0)$ and $\Sigma_1 = \text{syn}(X_1)$,
where $X_C$ is the sub-string of $X$ restricted to the indices in $\mathcal{I}_C$.
Similarly, one defines Bob's sub-strings $Y_0$ and $Y_1$.
The syndrome vectors are obtained in order to allow Bob to correct the errors in his local string.

\item Alice outputs two strings of $\ell$ bits, $S_0 = F_0(X_0)$ and $S_1 = F_1(X_1)$~\cite{fnote1}. 

\item Bob uses the syndrome $\Sigma_B$ to correct the errors in $Y_B$ and retrieve $\widetilde X_B$.
Finally, he applies the corresponding hash function and outputs $\widetilde S_B = F_B( \widetilde X_B)$.

\end{itemize}

\begin{remark}
Note the role of the waiting time $\Delta t$. 
If Bob is honest, he will measure as soon as he receives the photons.
If Bob is dishonest, and is storing the received qubits in a quantum memory, the waiting time will give some guarantees to Alice that Bob's quantum memory has at least partially decohered. 
\end{remark}

\begin{remark}
If the users are honest, then the protocol implements 1-2 $\text{ROT}^\ell$ correctly, up to a probability $\epsilon_\text{EC}$ that error correction fails.
Successful error correction means 
$\widetilde X_B = X_B$ and $\widetilde S_B = S_B$.
\end{remark}


\section{Security of the protocol}\label{Sec:sec}

The security analysis can be found in the literature \cite{OT,Noise-0} and is based on the following assumptions:
\begin{enumerate}
    \item Users have full knowledge and control of their own devices;
    \item Users have access to a noisy quantum memory. (The security analysis relies on modeling the noise in the quantum memory. Below we make use of a few explicit models);
    \item Loss and noise in the communication line and in the devices held by honest users are known publicly.
\end{enumerate}
Assumption 1 puts this protocol in the framework of device-dependent cryptography. Assumption 2 is the core assumption of the noisy-storage model. Assumption 3 is necessary to allow for error correction.
Furthermore, when analysing the security we need to consider that only one user is cheating (either Alice or Bob) and the other is honest, since
the OT protocol is designed to protect at least one honest user.
Receiver security is for Bob when Alice is cheating. 
Sender security is for Alice when Bob is cheating. 

Receiver security follows from the fact that physical qubits only travel from Alice to Bob. 
Note that there is some information flowing from Bob to Alice, due to the fact that Bob needs to confirm which photons have arrived. 
However, if Bob has full control over his device (from assumption 1) and behaves honestly (here we are considering the case where it is Alice who may cheat), this information cannot be used by Alice to guess the value of $B$. 
In fact, the data sent from Bob to Alice is only to confirm receipt of a photon, and do not convey any information about its internal degrees of freedom. Information about the state of the received photons may leak to Alice only if Bob's device is compromised.

Sender security relies on the noisy-storage assumption. %
The parameter $\ell$ depends on the amount of noise that affects dishonest Bob's quantum storage during the waiting time $\Delta t$. 
It also depends on the trusted noise, as some information will leak through the error syndromes.
Consider first the ideal case where error correction is not needed (i.e.~there is no noise for honest users), $\ell$ is estimated from the leftover hash lemma and expressed in terms of the smooth min-entropy~\cite{hash} (all logarithms are in base 2)
\begin{align}\label{OTminE}
    \ell \geq H_{\min}^{\epsilon_\text{s}} ( X_{1-B} | \mathcal{F}(Q) \Theta B ) 
    - 2 \log{\frac{1}{\epsilon_{\text{h}}}} + 1 \, ,
\end{align}
where $Q$ is the quantum information stored in the memory, and $\mathcal{F}$ is the map that describes the noisy storage for time $\Delta t$.
Such a value for $\ell$, if larger than zero, would ensure that the dishonest receiver cannot do much more than a random guess to determine the value of the complementary string $X_{1-B}$.
Quantitatively, the probability that the string remains unknown is given by the sum of the smoothing and hashing parameters, $\epsilon_\text{s} + \epsilon_\text{h}$.
However, noise also affects the honest users, therefore we need to employ error correction. In turn, this reduces the value of $\ell$, as a cheating receiver can in principle leverage the syndrome $\Sigma_{1-B}$ to acquire more information about the complementary bit string,
\begin{align}
    \ell & \geq H_{\min}^{\epsilon_\text{s}} ( X_{1-B} | \mathcal{F}(Q) \Theta B \Sigma_{1-B} ) - 2 \log{\frac{1}{\epsilon_{\text{h}}}} + 1 \\
         & \geq H_{\min}^{\epsilon_\text{s}} ( X_{1-B} | \mathcal{F}(Q) \Theta B ) - |\Sigma_{1-B}| 
         - 2 \log{\frac{1}{\epsilon_{\text{h}}}} + 1
         \, ,
         \label{ERCORR}
\end{align}
where in the second line we have applied a chain rule, and $|\Sigma_{1-B}|$ is the size of the syndrome in bits.
In principle, security is achieved whenever $\ell > 0$, for sufficiently small values of $\epsilon_\text{s} + \epsilon_\text{h}$.
Note that $\mathcal{F}(Q)$ is the quantum information in the noisy quantum memory, at the time where the basis information $\Theta^n$ is obtained by Bob.

Both the min-entropy and the length of the syndromes can be computed given suitable model for the noisy storage. This will be discussed in detail the next session.

\begin{remark}
Alice needs to run a statistical test to check that empirical attenuation factors are compatible with the expected value. 
Such a test is probabilistic and can fail with a probability $\epsilon_\text{test}$, which contributes to the security parameter of the protocol.
\end{remark}

In conclusion, if the protocol does not abort, it correctly implements 1-2 $\text{ROT}^\ell$ up to a failure probability $\epsilon_\text{EC}$. The implementation is secure against a dishonest receiver with noisy storage up to a probability $\epsilon_\text{s} + \epsilon_\text{h} + \epsilon_\text{test}$.


\section{Review of entropic bounds}\label{Sec:entropic}
In this Section we review a few \textit{entropic uncertainty relations} that have been used in the literature to prove the security of 
1-2 $\text{ROT}^\ell$~\cite{OT,Konig,ES}, in particular to establish sender security.
In fact, an entropic uncertainty relation can be applied to obtain a lower bound on the uncertainty of Bob in guessing the other string.

As a first step, consider a simplified scenario where Bob does not have any quantum memory. Therefore, he is forced to measure the quantum states as soon as he obtains them.
If the protocol passes the statistical test, we focus on the $n$ photons that have been tagged as received.
Alice has encoded the variables $X$ into their quantum states, using the BB84 encoding scheme with two conjugate bases.
After the waiting time $\Delta t$, Alice has announced her basis choices $\Theta$.
It is crucial, in this scenario, that Bob has already measured his quantum states when this basis information is revealed.
For example, this could happen because Bob has no quantum memory at all, or a quantum memory that is completely decohered after the waiting time $\Delta t$. In either case, Bob is expected to have already measured the states when he receives the basis information, that is, he has no quantum side information to rely upon.
In this scenario, the following entropic bound applies~\cite{OT},
\begin{align}\label{EUR0}
    H^\epsilon_{\min}(X | \Theta) \geq 
    \left(  \frac{1}{2} - 2\lambda \right)
    n  \, ,
\end{align}
where the parameter $\lambda$ can be chosen in the open interval $(0,1/2)$ such that
\begin{align}
    \epsilon = \exp{\left[  - \frac{\lambda^2 n}{32 ( \log(4/\lambda))} \right]} \, .
\end{align}

To make things more interesting, consider that a cheating receiver, when obtaining the basis information, still has $q = \nu n$ qubits stored in its quantum memory $Q$, which he has not measured yet.
In this case we have
\begin{align}
    H^{\epsilon}_{\min}(X | Q \Theta ) & \geq 
    H^{\epsilon}_{\min}(X | \Theta ) - q \\
    & \geq 
    \left(  \frac{1}{2} - 2\lambda \right)
    n - q 
    \\
    & =
    \left(  \frac{1}{2} - \nu - 2\lambda \right)
    n
    \, ,
\label{first_bound}
\end{align}
where the first inequality follows from the chain rule.
The parameter $\nu \in (0,1)$ represents the quantum storage rate.
Asymptotically in $n$, the bound in Eq.~(\ref{first_bound}) is non-trivial for $\nu < 1/2$, i.e., when a cheating receiver can store no more that one half of the qubits~\cite{fnote}.


\subsection{Noisy storage}\label{ssec:noisy}

In the case of noisy storage, the qubits stored in the quantum memory are partially degraded.
To analyse the security of OT in this scenario, we need to specify a model for the noisy quantum memory, described by the quantum channel $\mathcal{F}$.
Consider a quantum memory where each qubit is subject to independent and identically distributed (i.i.d.)~noise. 
For example, each qubit undergoes depolarising noise
\begin{align}\label{depol}
    \rho \to r \rho + (1-r) I/2 \, ,
\end{align}
which maps $\rho$ into the maximally mixed state with probability $1-r$, and leaves it untouched with probability $r$.

Note that a cheating receiver does not know if a particular qubit has been depolarised or not.
However, if we give him this additional information, we are making his quantum memory less noisy. 
In turn, this means that his uncertainty about the string can only decrease \cite{Konig,Thesis}. 
In average, if $q = \nu n$ are stored in the quantum memory, about $r \nu n$ of them are preserved without noise, whereas the remaining $(1-r) \nu n$ are completely depolarised.
Therefore, starting from (\ref{first_bound}), we obtain the following lower bound on the min-entropy (asymptotically in $n$):
\begin{align}\label{EURr0}
    H^{\epsilon}_{\min}( X | \mathcal{F}(Q) \Theta ) 
    & \geq 
    \left(  \frac{1}{2} - r \nu - 2\lambda \right)
    n 
    \, .
\end{align}
For large $n$, this is a non-trivial bound as long as $r \nu < 1/2$~\cite{fnote}.

The bound can be improved using the notion of \textit{strong converse} of a quantum channel for sending classical information \cite{Konig,sc}. 
Given that the noisy quantum memory is described by a map $\mathcal{F}$ applied to the qubits stored in the quantum memory, the ability of this channel to preserve (classical) information is quantified by 
\begin{align}
    P^\mathcal{F}_\text{succ}(n R )
    = \max_{\{\rho_x,D_x\}_x } \frac{1}{2^{nR}} \sum_{x \in \{0,1\}^{nR}} \mathrm{Tr}[ D_x \mathcal{F}(\rho_x)] \, ,
\end{align}
which is the maximum achievable (average) guessing probability, where the maximisation is over encoding states $\rho_x$ and decoding POVM $D_x$, given a bit-rate of $R$ bits per qubit.
The bound of K\"onig \textit{et al.}~\cite{Konig} reads
\begin{align}
H^{\epsilon+\epsilon'}_{\min}(X | \mathcal{F}(Q) \Theta ) 
 \geq 
    - \log{ P^\mathcal{F}_\text{succ}
    \left( 
    H^\epsilon_{\min}(X|\Theta) - \log{\frac{1}{\epsilon'}}
    \right) 
    }
     \, .
\end{align}

If the channel $\mathcal{F} = \mathcal{N}^{\otimes n}$ is i.i.d.,
then the entropic bound can be written explicitly for any $R > C_\mathcal{N}$, where $C_\mathcal{N}$ is the strong-converse capacity of the channel $\mathcal{N}$, and if the error exponent $\gamma(R)$ is known such that
\begin{align}
    P^{\mathcal{N}^{\otimes n}}_\text{succ} (nR) \leq 2^{- n \gamma(R)} \, .
\end{align}
Note that, if a strong converse exists, then $\gamma(R)>0$ for any $R>C_\mathcal{N}$. 

For $\mathcal{N}$ the depolarising channel in Eq.~(\ref{depol}), the strong-converse capacity is \cite{sc} \begin{align}
   C_\mathcal{N} =
   1 - h\left( \frac{1+r}{2} \right)
\end{align}
where 
\begin{align}
h(x) := - x \log{x} - (1-x)\log{(1-x)} 
\end{align}
is the binary Shannon entropy, 
and the error exponent is
\begin{align}
    \gamma_r(R) 
    & = 1 +
    \max_{\alpha > 1}
    \frac{ (\alpha-1)(R-1)
    - \log{\left[
    (1 + r)^\alpha + (1 -r)^\alpha
    \right]} }{\alpha} 
    \, .
\end{align}
This yields the entropy bound
\begin{align}
H^{\epsilon+\epsilon'}_{\min}( X | \mathcal{F}(Q) \Theta ) 
\geq 
    n \gamma_r \left( 
    \frac{ H^\epsilon_{\min}(X|\Theta ) - \log{(1/\epsilon')}
     }{n}
    \right) 
     \, .
\end{align}
Using Eq.~(\ref{EUR0}) we obtain
\begin{align}\label{EURr1}
H^{\epsilon+\epsilon'}_{\min}(X | \mathcal{F}(Q) \Theta ) \geq 
    n \gamma_r \left( 
  \frac{1}{2} - 2\lambda 
     - \frac{1}{n}\log{ \frac{1}{\epsilon'} }
    \right) 
     \, .
\end{align}
For $n$ sufficiently large, we obtain a bound on the asymptotic entropy rate
(note that this rate, as well as all the entropy rates computed in this paper, are expressed in bits per photon received)
\begin{align}\label{noisyK}
    h_{\mathrm{min}} 
    = \lim_{n\to\infty} \frac{1}{n} \, 
    H^{\epsilon+\epsilon'}_{\mathrm{min}}
    ( X | \mathcal{F}(Q) \Theta ) 
    \geq
    \gamma_r \left( 1/2 \right) 
     \, ,
\end{align}
which is a non-trivial bound for all values of $r$ such that $C_\mathcal{N} < 1/2$.

\begin{figure}[t!]
\includegraphics[width=0.9\linewidth]{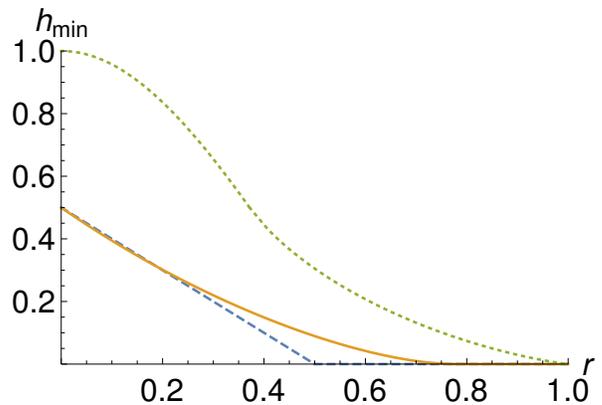}
\caption{Min-entropy rate vs the depolarising channel parameter.
Dashed blue line: computed from Eq.~(\ref{EURr0}) with $\nu = 1$.
Solid orange line: computed from  Eq.~(\ref{noisyK}).
Dotted green line: computed from  Eq.~(\ref{EURr2}).
}
\label{fig:ES}
\end{figure}

Figure~\ref{fig:ES} shows a comparison of the asymptotic entropy rate
$h_{\text{min}}$ for the depolarising channel, computed from Eq.~(\ref{EURr0}) with $\nu = 1$ (dashed blue line) and from Eq.~(\ref{noisyK}) (solid orange line).
This shows that Eq.~(\ref{noisyK}) is generally tighter for the depolarising-noise channel, but Eq.~(\ref{EURr0}) is tighter for small values of the depolarising parameter $r$, though with a relatively small gap.

For simplicity, for most of the rest of the paper we assume $\nu=1$. The general case of bounded storage ($\nu<1$) will be discussed in Section \ref{Sec:bounded}.

\subsection{Uncertainty relations for any amount of noise}

As we have seen for the depolarising channel, the entropic bounds we have obtained become trivial if the channel is not sufficiently noisy.
This goes against our physical intuition, which suggests that even a relatively weak noise may wipe at least some information.
In this Section we review an entropy bound, obtained by Dupuis \textit{et al.}~\cite{ES}, which is non-trivial even for low-noise quantum memory.

To write down this entropic bound explicitly, first consider a purification of the BB84-like protocol where Alice prepares $n$ copies of the maximally entangled two-qubit states $\Psi$, and sends to Bob one qubit from each pair.
To these states, which are stored in cheating Bob's quantum memory, a certain noisy channel is applied, yielding the $2n$-qubit state
\begin{align}
    \sigma_{AE} = \mathrm{id} \otimes \mathcal{F} (\Psi^{\otimes n}) \, ,
\end{align}
where $\mathrm{id}$ is the identity channel acting on the first qubit of each pair. 
The qubits on Alice's side are collectively indicated as $A$, and those stored in the quantum memory as $E = \mathcal{F}(Q)$.

The entropic bound is written in terms of the \textit{collision entropy rate} of such a state:
\begin{align}
    h_2(\sigma) := \frac{1}{n} \, H_2( A | E )_\sigma \, .
\end{align}
Recall that the collision entropy of the bipartite state $\sigma$ is defined as
\begin{align}
    H_2( A | E )_\sigma & = 
    - \log{ \mathrm{Tr}\left[ \left( \sigma_E^{-1/4} \sigma_{AE} \, \sigma_E^{-1/4} \right)^2 \right]} 
    \, ,
\end{align}
where $\sigma_E = \mathrm{Tr}_A \sigma_{AE}$ is the reduced state obtained by partial tracing.

The following min-entropy bound is proven in Ref.~\cite{ES}: 
\begin{align}\label{ESeq}
    H_\mathrm{min}^\epsilon( X | E \Theta ) \geq n \Gamma(h_2(\sigma)) - 1 - \log{\frac{2}{\epsilon^2}} \, ,
\end{align}
where the function $\Gamma$ is defined as
\begin{align}
    \Gamma(x) = \left\{
    \begin{array}{ccc}
    x         & \mbox{if} & x \geq 1/2 \, , \\
    g^{-1}(x) & \mbox{if} & x < 1/2 \, .
    \end{array}
    \right.
\end{align}
and
\begin{align}
    g(y) = - y \log{y} - (1-y) \log{(1-y)} + y - 1 \, . 
\end{align}

If the noisy quantum memory is described by an i.i.d.~depolarising channel, then the state $\sigma$ is a direct product, i.e.~$\sigma = \tau^{\otimes n}$, with 
\begin{align}\label{eq:tau}
    \tau = r \Psi + (1-r) I/2 \otimes I/2 \, ,
\end{align}
and the collision entropy reads
\begin{align}\label{h2sm1}
    h_2(\sigma) 
    & = - \log{2 \mathrm{Tr}(\tau^2)}
    = 1 - \log{ ( 1 + 3r^2 ) } \, .
\end{align}
This yields
\begin{align}
    H_\mathrm{min}^\epsilon( X | \mathcal{F}(Q) \Theta ) \geq n \Gamma[1 - \log{ ( 1 + 3r^2 ) }] - 1 - \log{\frac{2}{\epsilon^2}} \, .
\end{align}
Note that, at least for $n$ large enough, this bound remains non-trivial even when $r$ is arbitrarily close to $1$, with the asymptotic entropy rate
\begin{align}\label{EURr2}
    h_{\text{min}} 
    \geq \Gamma[1 - \log{ ( 1 + 3r^2 ) }]  \, .
\end{align}
This is plotted in Figure \ref{fig:ES} (dotted green line), showing that this latter bound supersedes those obtained in the previous Sections.


\section{Improved min-entropy bounds}\label{Sec:improved}

In this Section we derive a new min-entropy bound using the uncertainty relation of Ref.~\cite{ES}.
The argument is analogous to the one used in Section~\ref{ssec:noisy} to obtain Eq.~(\ref{EURr0}).
The difference is that~(\ref{EURr0}) was obtained from the uncertainty relation~(\ref{EUR0}), whereas here our starting point is the uncertainty relation in Eq.~(\ref{ESeq}).
To obtain this bound, we assume that each qubit received by Bob is affected by identical and independent noise. As above, this qubit noise is modeled as a depolarising channel.

Consider the depolarising channel of Eq.~(\ref{eq:tau}), which preserves the state with probability $r$ and completely depolarises it with probability $1-r$. Cheating Bob does not know whether a given qubit has been depolarised or not while stored in his quantum memory. 
However, if we give him this additional information, the depolarising channel is replaced by the erasure channel:
\begin{align}
    \tau' = r \Psi + (1-r) I/2 \otimes \omega \, ,
\end{align}
where $\omega$ is the erasure flag, which allows Bob to know if the state has been stored without error by applying a non-destructive measurement.
Given that $n_1$ qubits have been preserved without error, and $n-n_1$ have been erased, the overall $n$-qubit state reads
\begin{align}
    \sigma' = \Psi^{ \otimes n_1 } \otimes (I/2 \otimes \omega)^{\otimes (n-n_1)} \, .
\end{align}
Assume Bob is given knowledge of which qubits have been erased. Denote by $X^{n_1}$ the sub-string of bits corresponding to the qubit that have been preserved without noise, and by $X^{n-n_1}$ the substring 
corresponding to the qubits that have been erased, with $X = X^{n_1} X^{n-n_1}$.
We can then write a lower bound on the min-entropy:
\begin{align}
H_{\min}^\epsilon (X | E \Theta ) 
& \geq H_{\min}^\epsilon (X^{n-n_1} | E \Theta ) \\
& \geq (n-n_1) \Gamma(h_2(\sigma'')) - 1 - \log{\frac{2}{\epsilon^2}} \, ,
\end{align}
where the first inequality comes from the fact that the entropy of a bit string is always larger than the entropy of a substring, the second inequality is an application of Eq.~(\ref{ESeq}), and $\sigma'' = (I/2 \otimes \omega)^{\otimes (n-n_1)}$.
Noting that $h_2( I/2 \otimes \omega ) = 1$ and $\Gamma( 1 ) = 1$, we obtain
\begin{align}
H_{\min}^\epsilon (X | E \Theta ) 
\geq n-n_1 - 1 - \log{\frac{2}{\epsilon^2}} \, .
\end{align}

\begin{figure}[t!]
\includegraphics[width=0.9\linewidth]{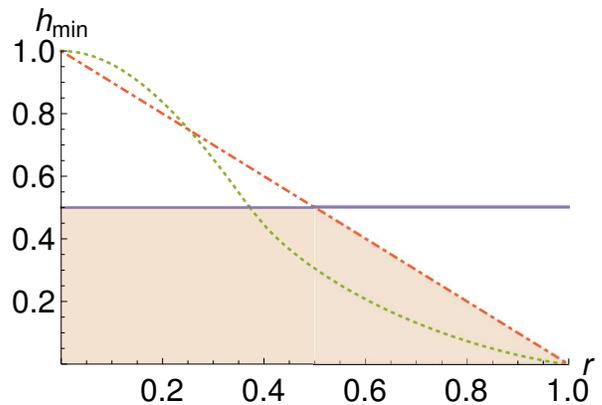}
\caption{Min-entropy rate vs the depolarising noise parameter.
Dotted green line: computed from Eq.~(\ref{EURr2}) (same as shown in Fig.~\ref{fig:ES}).
Dash-dotted red line: computed from Eq.~(\ref{linear}).
The best entropy rate for each value of $r$ is obtained by taking the maximum of the two curves, as in Eq.~(\ref{eq:max}).
Solid blue line: min-entropy for the honest receiver, obtained from Eq.~(\ref{EUR0}) in the limit of large $n$.
If the receiver behaves rationally, his entropy rate is always above the shadowed region.}
\label{fig:linear}
\end{figure}

In the limit of large $n$, the number of virtually erased qubits is expected to be $n-n_1 \simeq (1-r) n$. Therefore we obtain the bound on the asymptotic min-entropy rate
\begin{align}\label{linear}
    h_{\min} \geq 1-r  \, .
\end{align}
This new bound is shown together with the previous one in Fig.~\ref{fig:linear}.
In conclusion, for the depolarising channel, the best bound on the entropic rate obtained so far is
\begin{align}\label{eq:max}
    h_{\min} \geq \max\left\{ \gamma[1 - \log{ ( 1 + 3r^2 ) }] , 1-r \right\} \, .
\end{align}
As shown in the figure, for smaller values or $r$ (noisier quantum memory) the best bound is $h_{\min} = \gamma[1 - \log{ ( 1 + 3r^2 ) }]$, whereas for higher values of $r$ (less noisy quantum memory) the best bound is $h_{\min} = 1-r$. 

\begin{figure*}[t!]
\includegraphics[width=0.9\linewidth]{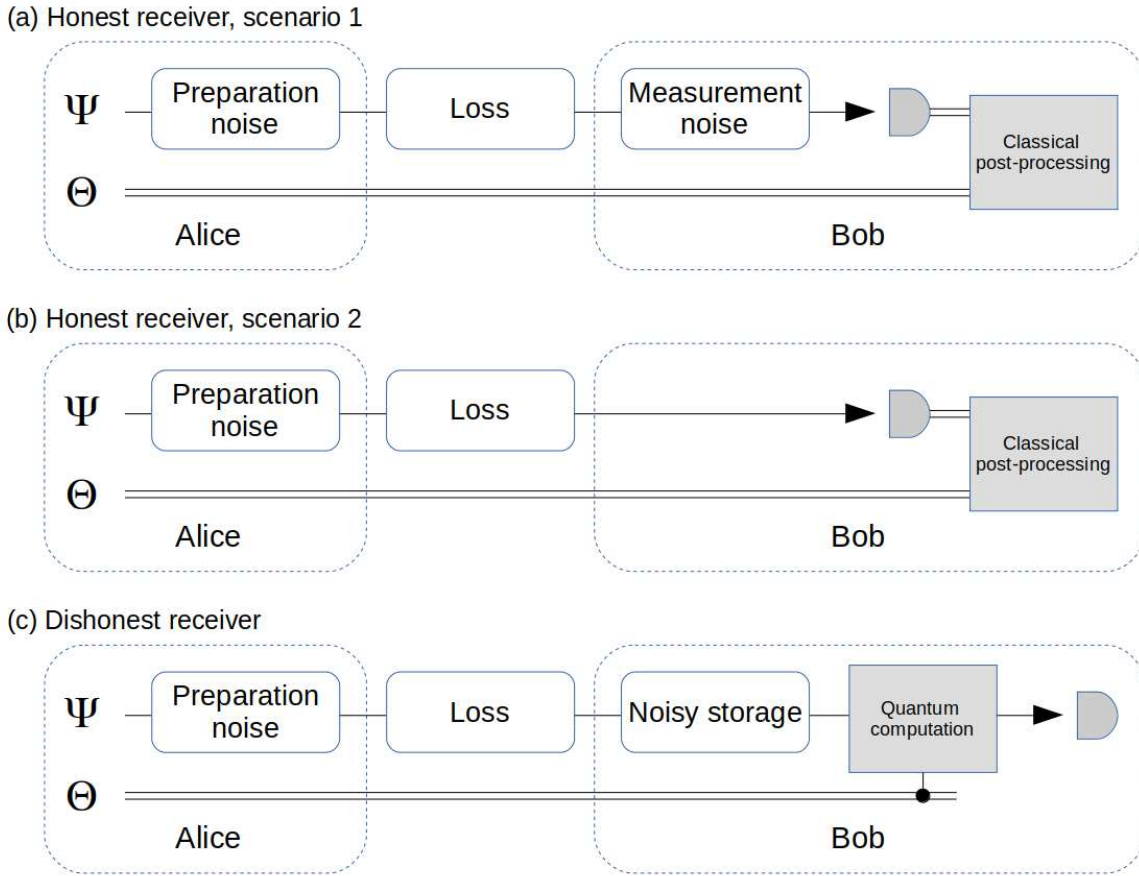}
\caption{Preparation noise is associated to the sender (Alice) station.
Losses, including those occurring during the preparation phase and due to detector inefficiencies, are associated to the communication channel between Alice and Bob.
(a) In scenario 1, honest receiver Bob measures the quantum states has soon as he gets them, using a noisy measurement apparatus.
(b) In scenario 2, the honest receiver's measurement apparatus has negligible noise.
(c) A cheating receiver holds the quantum states in a noisy quantum memory until he receives the basis information from Alice.
}
\label{fig:diagrams}
\end{figure*}


\subsection{Optimal strategies for a dishonest receiver}

The entropic bounds obtained so far allow us to estimate the uncertainty of the dishonest receiver in guessing Alice's string, as a function of the noise affecting the quantum memory.
Recall that this is the noise describing the state of the qubits stored in the quantum memory when Bob obtains the basis information $\Theta$.
We suppose that Bob is not acting honestly, as according to the protocol he should measure the qubits as soon as he receive them.
Equation (\ref{eq:max}) shows that the uncertainty increases with increasing noise in the quantum memory. 
Eventually, if the memory is too noisy, crime does not pay anymore, 
and being dishonest (i.e.~Bob waiting for the basis announcement before measuring) is no longer
the rational choice.

In fact, we know from Eq.~(\ref{EUR0}) that the min-entropy rate for the honest receiver is $h_{\min} \geq 1/2$.
Therefore, dishonest behaviour is no longer rational if the value on the right-hand-side of Eq.~(\ref{eq:max}) is larger than $1/2$.
When the receiver acts rationally and applies the best strategy, the entropy rate is 
\begin{align}
    h_{\min} & \geq \min\left\{ 1/2 , \max\left\{ \gamma[1 - \log{ ( 1 + 3r^2 ) }] , 1-r \right\} \right\} \\
    & = \min\left\{ 1/2 , 1-r \right\} \, .
    \label{rational}
\end{align}
In conclusion, for a rational receiver the entropy rate is always above the shadowed region in Fig.~\ref{fig:linear}.

This result is indeed intuitive.
Since the honest receiver can measure without error about $50\%$ of the qubits, behaving honestly is the rational choice whenever the quantum memory corrupts more that $50\%$ of the qubits.
Therefore, we expect the bound (\ref{rational}) to be tight if the memory is modeled as an erasure channel (with a flag), where $1-r$ is the probability of erasing the qubit.
However, for the depolarising channel this bound is not expected to be tight and there might be room for improvement.

\begin{figure}[t!]
\centering
\subfigure{\includegraphics[width=0.4\textwidth]{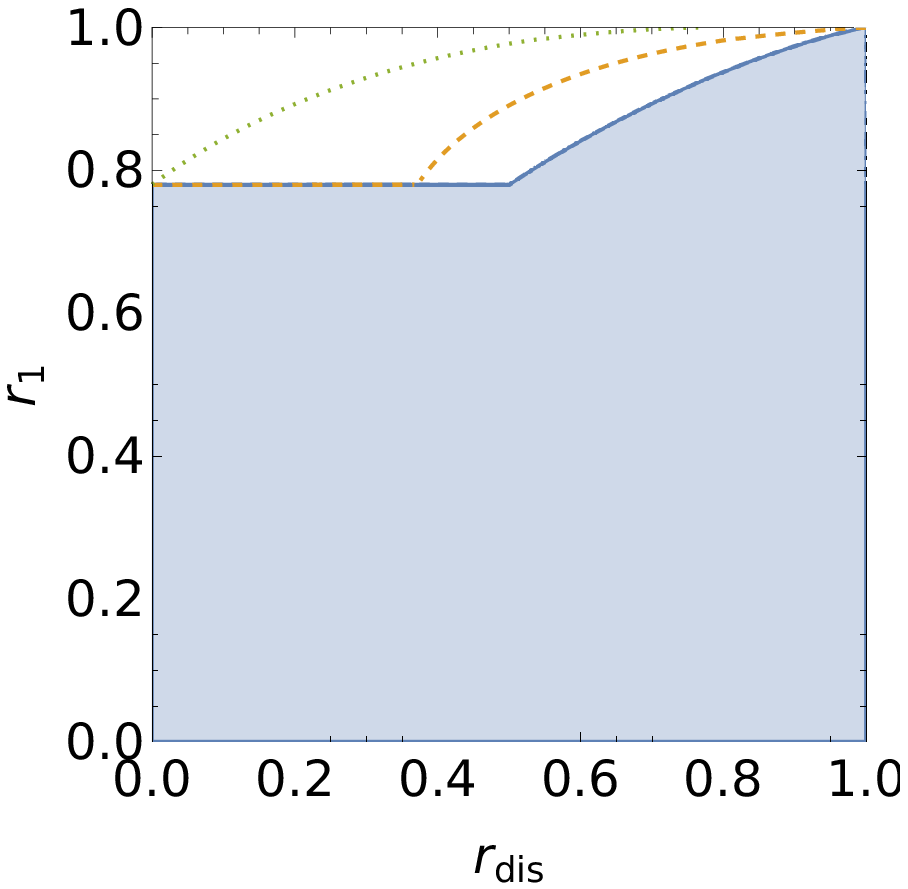}} \\
\subfigure{\includegraphics[width=0.4\textwidth]{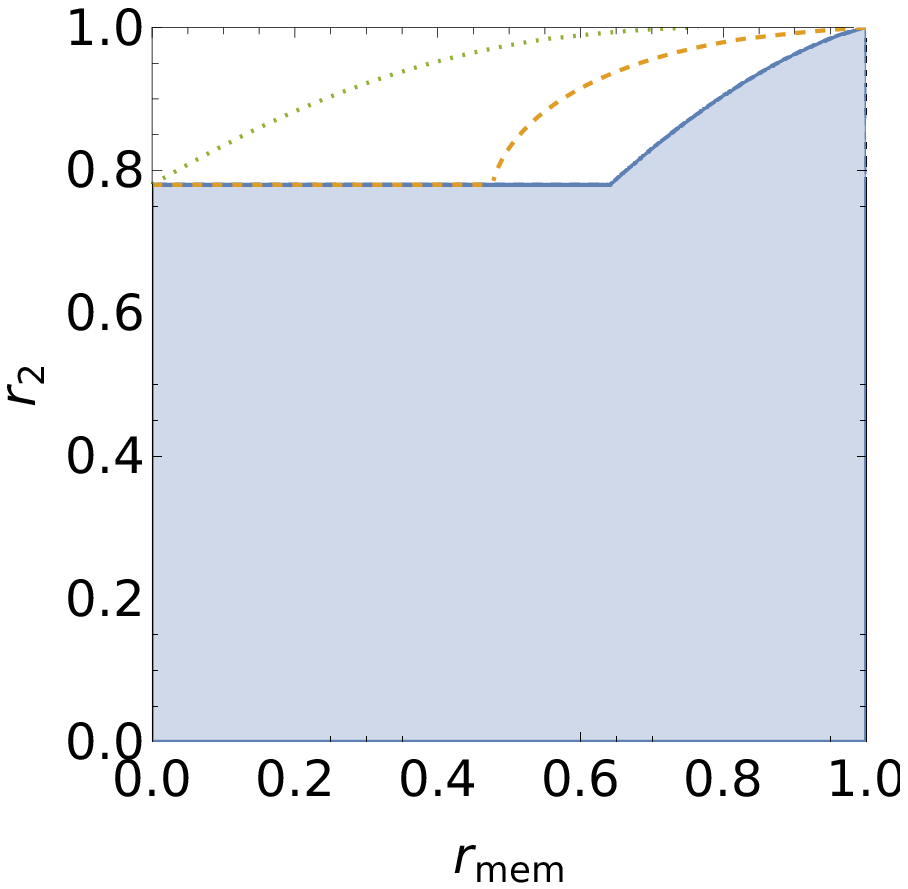}}
\caption{
Top: scenario 1. Secure OT can be achieved above the shadowed region in the $r_\text{dis}$-$r_1$ plane. 
Bottom: scenario 2. Secure OT can be achieved above the shadowed region in the $r_\text{mem}$-$r_2$ plane.
For comparison with previous results from \cite{Konig,ES}, the dotted green line shows the boundary between secure and non-secure regions that would be obtained using the min-entropy bound of Eq.~(\ref{noisyK}) [see Eqs.~(\ref{noisyKrat}) and~(\ref{noisyKrat_2})].
The orange dashed line is the boundary that would be obtained from the min-entropy rate in Eq.~(\ref{EURr2}) [see Eqs.~(\ref{EURr2rat}) and~(\ref{EURr2rat_2})]. 
}
\label{fig:scen12}
\end{figure}


\section{Error-tolerant OT}\label{Sec:tolerant}

The entropic uncertainty relations of the previous Sections can be directly applied to bound the uncertainty of the receiver Bob about the string $X$.
Note, however, that for application to OT, we are interested in bounding the uncertainty of a cheating receiver, not about the whole string $X$, but only about the substring $X_{1-B}$. Therefore we need a lower bound on the min-entropy of $X_{1-B}$, as in Eq.~(\ref{OTminE}).
For large enough $n$, the substring $X_{1-B}$ is expected to have size of about $n/2$ bits, which are randomly sampled from $X$. 
According to the min-entropy sampling discussed in Ref.~\cite{Konig}, the min-entropy rate of a randomly chosen substring is the same as $X$, if $n$ is large enough, up to finite-size corrections. Therefore, in the asymptotic limit of large $n$ we can use the min-entropy rates as obtained in the previous Sections.
To account for noise in their devices, Alice and Bob will also apply error correction, which further reduces the min-entropy as in Eq.~(\ref{ERCORR}).

We consider two scenarios, depicted in Fig.~\ref{fig:diagrams}(a) and~\ref{fig:diagrams}(b).
In both scenarios, Alice's device introduces some noise during the phase of state preparation, and photon loss occurs in the transmission from Alice to Bob.
In scenario 1, Bob's device is noisy and lossy, for example due to non-unit detector efficiency.
In scenario 2, Bob's devices is noiseless, and only affected by loss.
In both cases, to simplify the analysis, we model all noises (in preparation and measurement) as depolarising noise.
Also, loss during state preparation and measurement are (with no loss of generality) associated to the communication channel. 
This is consistent as loss commutes with depolarising noise.
Let $\eta$ indicate the total attenuation factor, accounting for loss in state preparation, transmission, and measurement (non-unit detection efficiency).
As discussed above, to deal with loss, Alice needs to know the expected value of $\eta$, and she will abort the protocol if the empirical value is too different from the expected one given the statistical fluctuations.

In scenario 1, the total noise is obtained by combining two depolarising channels (modeling preparation noise and measurement noise), yielding to an overall depolarising channel with parameter
\begin{align}
    r_1 = r_\text{pre} r_\text{mea} \, . 
\end{align}
Scenario 2 is less noisy, and depolarisation is only due to the preparation phase,
\begin{align}
    r_2 = r_\text{pre} \, . 
\end{align}
These relations hold for the honest receiver in the two scenarios.
If the receiver is dishonest and stores the qubits in a quantum memory, from his point of view the noise in the quantum memory combines with the noise in the preparation phase, as shown in Fig.~\ref{fig:diagrams}(c).
As above, we model the noisy storage as a depolarising channel with parameter $r_\text{mem}$.
This means that the dishonest receiver will experience a total depolarising noise with parameter
\begin{align}
    r_\text{dis} = r_\text{pre} r_\text{mem} 
\, . 
\end{align}
Note that $r_\text{dis} = r_2 r_\text{mem} \leq r_2$, and
scenario 2 is more advantageous for the honest users, as the noise experienced by the cheating receiver is always larger than the honest one.

For each scenario, the amount of error correcting information per channel use is asymptotically equal to
\begin{align}
    h_\text{EC} = h\left( \frac{1+r_j}{2} \right) \, ,
\end{align}
for $j=1,2$ according to the scenario considered.
From this we can compute the asymptotic communication rates, measured in bit per channel use, using Eq.~(\ref{ERCORR}).
For scenario 1 we obtain
\begin{align}
b :=    \lim_{n\to\infty} \ell/n = 
    \min\left\{ 1/2 , 1 - r_\text{dis} \right\} 
    - h\left( \frac{1+r_1}{2} \right) \, .
\end{align}
Figure~\ref{fig:scen12} (top) shows the contour plot of the bit rate.
The protocol is secure for values of $r_\text{dis}$ and $r_1$
above the shadowed region.
The figure also shows how this result improves on existing literature. 
The green dotted line 
and the orange dashed line are the boundaries between the regions of secure and non-secure OT obtained from Eq.~(\ref{noisyK}) and Eq.~(\ref{EURr2}), respectively. 
In particular, for a rational receiver we obtain from Eq.~(\ref{noisyK}) the bit rate
\begin{align}\label{noisyKrat}
    \min\left\{ 1/2 , 
    \gamma_{r_\text{dis}} \left( 1/2 \right) \right\} 
    - h\left( \frac{1+r_1}{2} \right) \, .
\end{align}
Similarly, from Eq.~(\ref{EURr2}) we obtain
\begin{align}\label{EURr2rat}
    \min\left\{ 1/2 , 
    \Gamma[1 - \log{ ( 1 + 3r_\text{dis}^2 ) }] \right\} 
    - h\left( \frac{1+r_1}{2} \right) \, .
\end{align}

For scenario 2, we can use $r_\text{mem}$ as an independent variable, and the asymptotic rate is
\begin{align}
    b
    = 
    \min\left\{ 1/2 , 1 - r_ 2 r_\text{mem} \right\} 
    - h\left( \frac{1+r_2}{2} \right) \, .
\end{align}
The contour plot for this bound is shown in Fig.~\ref{fig:scen12} (bottom), where secure OT is achieved above the shadowed region in the $r_\text{mem}$-$r_2$ plane. 
The figure also shows the boundary between regions corresponding to secure and non-secure OT that would be obtained using the min-entropy bounds in Eq.~(\ref{noisyK}) and~(\ref{EURr2}).
The latter are computed from the bit rates
\begin{align}
    & \min\left\{ 1/2 , 
    \gamma_{r_ 2 r_\text{mem}} \left( 1/2 \right) \right\} 
    - h\left( \frac{1+r_2}{2} \right) \, ,
    \label{noisyKrat_2}\\
    & \min\left\{ 1/2 , 
    \Gamma[1 - \log{ ( 1 + 3(r_ 2 r_\text{mem})^2 ) }] \right\} 
    - h\left( \frac{1+r_2}{2} \right) \, .
    \label{EURr2rat_2}
\end{align}

Note that scenario 2 is always more advantageous for the honest users, but in both cases secure OT is possible only if the trusted noise parameter $r_j$ is such that $h((1+r_j)/2)\leq 1/2$, i.e.~$r_j \gtrsim 0.78$.
This corresponds to a maximum tolerable trusted noise of about 22\%.


\subsection{Bounded storage}\label{Sec:bounded}

In case of noisy and bounded storage, suppose that the cheating receiver can at most store a fraction of the received qubits, quantified by the quantum storage rate $\nu$. 
Equation~(\ref{eq:max}) is thus replaced by
\begin{align}\label{eq:maxB}
    h_{\min} 
    & \geq \frac{1-\nu}{2} + 
    \nu \max\left\{ \gamma[1 - \log{ ( 1 + 3r^2 ) }] , 1-r \right\} \, .
\end{align}
For a rational receiver, Eq.~(\ref{rational}) is replaced by
\begin{align}
    h_{\min} 
    & \geq \frac{1-\nu}{2}  \\
    & \phantom{=}~ + 
    \nu \min\left\{ 1/2 , \max\left\{ \gamma[1 - \log{ ( 1 + 3r^2 ) }] , 1-r \right\} \right\} \nonumber \\
    & = \frac{1-\nu}{2} + 
    \nu ( 1-r ) 
    = 1/2 + \nu(1/2 - r) \, .
    \label{rationalB}
\end{align}
Finally, the asymptotic rate for scenario 1 becomes
\begin{align}
    b = 
    \min\left\{ 1/2 , 1/2 + \nu(1/2 - r_\text{dis}) \right\} 
    - h\left( \frac{1+r_1}{2} \right) \, .
\end{align}
Similarly for scenario 2 we obtain
\begin{align}
    b = 
    \min\left\{ 1/2 , 
    1/2 + \nu(1/2 - r_ 2 r_\text{mem})
    \right\} 
    - h\left( \frac{1+r_2}{2} \right) \, .
\end{align}

\begin{figure}[t!]
\includegraphics[width=0.8\linewidth]{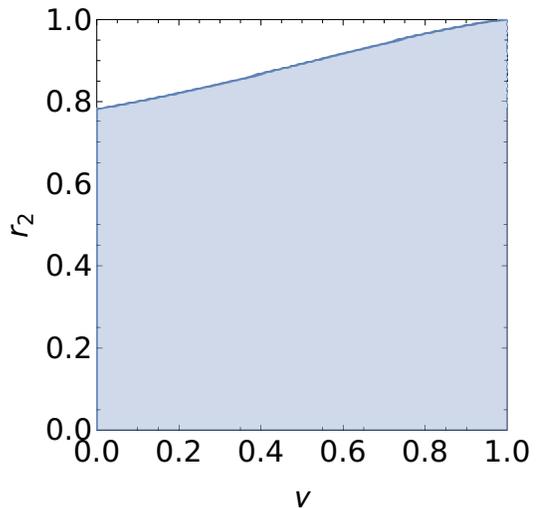}
\caption{Scenario 2, for $r_\text{mem}=0$. This refers to a noiseless yet bounded quantum memory, with storage rate $\nu$. The shadowed region corresponds to where the protocol is not secure in the $\nu$--$r_2$ plane, where $r_2$ is the trusted noise parameter.}
\label{fig:bounded}
\end{figure}

Let us consider in more detail scenario 2. In the case of noiseless but bounded quantum memory, we put $r_\text{mem}=1$ and the bit rate becomes
\begin{align}
b =
    \min\left\{ 1/2 , 
    1/2 + \nu(1/2 - r_ 2)
    \right\} 
    - h\left( \frac{1+r_2}{2} \right) \, .
\end{align}
Figure \ref{fig:bounded} shows the region where the rate vanishes in the $\nu$--$r_2$ plane. For $\nu$ approaching zero we recover the threshold of $22\%$ trusted noise. This threshold value decreases nearly linearly with increasing $\nu$.

\section{Correlated noise}\label{Sec:corr}

\begin{figure}[t!]
\includegraphics[width=0.9\linewidth]{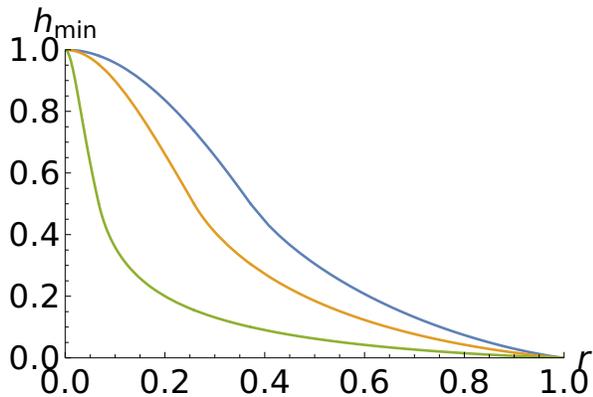}
\caption{Min-entropy bound in Eq.~(\ref{ESeq_corner}) plotted vs the noise parameter $r$. Shown for different values of the correlation parameter $m$. From top to bottom, the curves are obtained for $m=1,2,5$.}
\label{fig:corre}
\end{figure}

In this Section we discuss the case of quantum memory affected by correlated noise~\cite{RMP}. 
We consider a model of burst errors where $m > 1$ neighbour qubits are collectively depolarised.
The integer $m$ plays the role of a correlation parameter.
Given $m$ copies of the maximally entangled two-qubit states $\Psi$, this model is represented by the map
\begin{align}
\Psi^{\otimes m} \to 
\sigma_{AE} = r \Psi^{\otimes m}
+ 2^{-2m} (1-r) I^{\otimes 2m} \, ,
\end{align}
which replaces (\ref{eq:tau}).
To compute the bound in Eq.~(\ref{ESeq}) we first need to compute the collision entropy of the state $\sigma_{AE}$;
%
%
we obtain
\begin{align}
h_2 
= 1 - \frac{1}{m} \log\left[ 1 +  (2^{2m}-1) r^2 \right] \, ,
\end{align}
which extends Eq.~(\ref{h2sm1}) to any $m>1$.
From this we obtain the min-entropy rate for the correlated-noise quantum memory:
\begin{align}\label{ESeq_corner}
    h_\mathrm{min}
\geq \Gamma\left[ 
1 - \frac{1}{m} \log\left[ 1 +  (2^{2m}-1) r^2 \right]
\right]  
    \, .
\end{align}
As shown in Fig.~\ref{fig:corre} this bound decreases with increasing $m$.

As we did in Section \ref{Sec:improved}, we now improve this bound by introducing a virtual erasure channel. This time the erasure channel acts on $m$ qubits.
A collection of $m$ neighbour qubits is erased with probability $(1-r)$. If erased, we obtain the erasure flag $\omega$. To see the action on the whole set of $n$ qubits, we may split them in groups of $m$ neighbours. Each group is erased with probability $(1-r)$. Overall, for large $n$, we expect about $n_e = (1-r) n$ qubits to be erased. This corresponds to the $n$-qubit state
\begin{align}
    \sigma' & = \Psi^{ \otimes (n-n_e) } \otimes (I/2 \otimes \omega)^{\otimes n_e} \\
    & = \Psi^{ \otimes r n} \otimes (I/2 \otimes \omega)^{\otimes (1-r)n} \, .
\end{align}
From this state we compute the min-entropy rate
\begin{align}\label{linear2}
    h_{\min} \geq 1-r  \, ,
\end{align}
which is independent of $m$ and equal to the bound obtained in the case of i.i.d.~noise.

In conclusion, our new bound is always tighter, and the gap with the previous bound increases with increasing value of the correlation parameter $m$.


\section{Conclusions}\label{Sec:end}

Quantum mechanics allows for information-theoretically secure two-party computation. Unlike quantum key distribution, however, two-party computation is not unconditionally secure, but requires additional assumptions on the capability of a cheating party. In particular, one can achieve provably secure OT from Alice to Bob if the receiver Bob is limited in the amount or quality of his quantum memory, known as the noisy-storage model of quantum cryptography.

Experimental implementations of OT are particularly challenging, and much more challenging than QKD, because in two-party computation the users do not trust each other. This requires attention because Alice and Bob cannot cooperate, as they would do in QKD, in order to estimate the noise in the communication channel and apply error correction. 
In two-party computation, the honest users need to know in advance the noise and loss characteristics of the communication channel and of their trusted devices.

Intuitively, the more noisy the devices of the honest users are, the easier it is for a cheating user to hide in this trusted noise. This induces a trade-off between trusted noise (in the devices of the trusted users) and untrusted noise (in the quantum memory of a cheating receiver). This trade-off is ultimately quantified by the uncertainty relation used to assess the security of the OT protocol.

In this paper we have introduced improved entropic uncertainty relations and applied them to characterise the trade-off between trusted noise and quantum memory noise. 
We have also noted that cheating is not a rational behaviour if the quantum memory is too noisy.
We have shown that, even if the quantum memory is arbitrarily noisy, yet unbounded, the trusted noise, modeled as depolarising noise, cannot surpass $22\%$.
For low-noise quantum memory, secure OT can be achieved only if the trusted noise is also low, with an improved trade-off as shown in Fig.~\ref{fig:scen12}, or in the case of limited storage.

We have discussed depolarising noise but our results directly apply to a more general noise model of the form 
\begin{align}
    \rho \to r \rho + (1-r) \rho_0 \, ,
\end{align}
where $\rho_0$ is a fixed point independent of $\rho$.
For simplicity and clarity of exposition, here we have focused on the asymptotic limit of many channel uses.
However, our approach can be applied to the finite-size regime as well.


\begin{acknowledgments}
This work has received funding via the EPSRC Quantum Communication Hub (EP/T001011/1)'s partnership resource scheme and from
the European Union’s Horizon Europe research and innovation programme under the project "Quantum Secure Networks Partnership" (QSNP, grant agreement No 101114043).
C.L. acknowledges financial support from PNRR MUR project PE0000023-NQSTI.
\end{acknowledgments}



\begin{thebibliography}{99}
\bibitem{fnote1}
Note that $X_0$ and $X_1$ are shorter than $n$ bits, whereas the hash functions are from $n$ to $\ell$ bits. The notation $F_C(X_C)$ implicitly implies that a sufficient number of $0$'s are appended to $X_C$.
\end{thebibliography}
\end{document}